\documentclass[aps,twocolumn,showpacs,preprintnumbers,amsmath,amssymb]{revtex4}
\usepackage{graphicx}
\begin{document}

\title {Monte Carlo Simulations of Ultrathin Magnetic Dots}
\author{M. Rapini}
\affiliation{Laborat\'orio de Simula\c{c}\~ao - Departamento de
F\'{\i}sica - ICEX - UFMG 30123-970 Belo Horizonte - MG, Brazil}
\author{R.A. Dias}
\affiliation{Laborat\'orio de Simula\c{c}\~ao - Departamento de
F\'{\i}sica - ICEX - UFMG 30123-970 Belo Horizonte - MG, Brazil}
\author{D.P. Landau}
\affiliation{Center for Simulational Physics, University of Georgia,
Athens, Georgia 30602}
\author{B.V. Costa}
\affiliation{Laborat\'orio de Simula\c{c}\~ao - Departamento de
F\'{\i}sica - ICEX - UFMG 30123-970 Belo Horizonte - MG, Brazil}
\begin{abstract}
In this work we study the thermodynamic properties of ultrathin
ferromagnetic dots using Monte Carlo simulations. We investigate the
vortex density as a function of the temperature and the vortex
structure in monolayer dots with perpendicular anisotropy and
long-range dipole interaction. The interplay between these two terms
in the hamiltonian leads to an interesting behavior of the
thermodynamic quantities as well as the vortex density.
\end{abstract}

\maketitle

\section{Introduction}

Magnetism at nanoscale, when the size of the structure is comparable
to or smaller than both the ferromagnetic (FM) and antiferromagnetic
(AF) domain size, offers a great potential for new physics. In the
last decade there has been an increasing interest in ultrathin
magnetic dots from research groups as well as technological
industries. Such an interest is due to numerous unique phenomena
related to the low-dimension of these systems.

The modern technology demands techniques capable of producing
nanometer-sized structures over large areas. A good perspective is
the use of nanodots of nickel that could store terabyte of data in a
computer chip just a few centimeters wide. In particular,
ferromagnetic nanodots have been widely studied by use of
experimental techniques such as MFM (magnetic force microscopy). In
addition, some theoretical models were proposed to explain the
physical phenomena observed in the experiments, among them the
transition from perpendicular to in-plane ordering and the
magnetoresistence effect.

Regarding the perpendicular to in-plane ordering transition,
experiments were done using epitaxial films to investigate its
transition temperature and thickness dependence \cite{bischof}
\cite{pappas}. In addition, many theoretical approaches were
developed, for example, treating a two-dimensional layer by
renormalization group \cite{pescia}. Some lattice models were
proposed to take into account long-range dipolar interactions and
surface anisotropy \cite{moschel}.

Based on such models, Monte Carlo simulations have been widely used
to study the phase diagram of very thin films \cite{santamaria}, the
nature of this transition \cite{hucht} as well as its dependence on
the magnetic history of the system \cite{iglesias}. On the other
hand, magnetic domains \cite{matsubara} and magnetic structures
\cite{vedmedenko} have also been investigated by using computational
methods. A topological excitation, the spin vortex, has been found
in experiments and also detected in simulations. Vortex structures
are believed to drive a Bereziinski-Kosterlitz-Thouless (BKT) phase
transition in the two dimensional planar-rotator (PR) model
\cite{kogut}. Although vortices are present in thin films with long
range interactions, it is not clear if they play any role in the
transition.

The model we study is described by the Heisenberg spin hamiltonian
with exchange and long-range dipolar interactions as well as
single-ion anisotropy
\begin{widetext}
\begin{equation}
H= -J\sum_{<ij>}\bf S_{i} \cdot S_{j} \rm+ D \sum_{i \neq k}
\frac{\bf S_{i} \cdot
 S_{k}}{\rm r_{ik}^{3}}
-3\frac{(\bf S_{i} \cdot r_{ik})(S_{k} \cdot r_{ik})}{\rm
r_{ik}^{5}} -A\sum_{i}(S_{i}^{z})^{2} ~~~,
\end{equation}
\end{widetext}
where we use classical spins $|\bf S \rm |=1$. Here the first sum is
performed over nearest neighbors with exchange coupling strenght,
$J>0$ , while the second sum runs over all spin pairs in the
lattice. The constant of dipole coupling is $D$, ${\bf r}_{ik} $ is
a vector connecting the $i$ and $k$ sites and $A$ is the single-site
anisotropy constant along the z-axis\cite{moschel}.

The main task in this work is to study the importance of vortices in
the physics of the model. Although preliminary, our results indicate
an anomalous behavior of the vortex density at the transition
temperature for $\delta=\frac{D}{A} \ll 1$. In the following we
present a brief background on the simulation, our results and the
conclusions.
\\
\\
{\sl Method}
\\
\\
The simulations are done in a square lattice of volume $L \times L$
with $L=20,40,60$ by using the Monte-Carlo method with the
Metropolis algorithm \cite{mc1,mc2}. Since nanodots are finite {\it
per nature} we have to use open boundary conditions in our
simulations. However, we want to emphasize the long range effects of
the dipolar term of the model at the boundary of the structure. For
that, we have used periodic boundary conditions in the non dipolar
terms while for the dipolar term we have used open conditions.

We have studied the model for three different values of the
parameters $A$ and $D$, $\delta = \frac{D}{A}= 0.1, 1.0$ and $9.0$
for fixed $J=1$. Energy is measured in units of $JS^2$ and
temperature in units of $JS^2/k_B$, where $k_B$ is the Boltzman
constant. For every temperature the first $10^5$ MC steps per spin
were used to lead the system to equilibrium and the next $10^5$
configurations were used to calculate thermal averages of
thermodynamical quantities of interest.

\section{Results}

\begin {figure}[htbp]
\vspace{0.5cm}
\includegraphics [height=6.0cm,width=6.5cm] {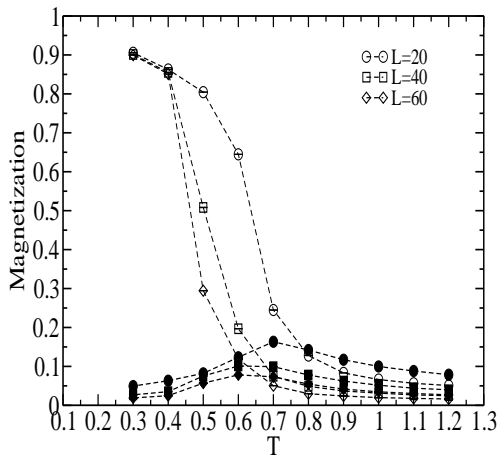}
\caption{Out-of-plane and in-plane magnetization (open and full
symbols) for $\delta =0.1$} \label{fig1}
\end {figure}

In the case where $\delta = 0.1$, we measured the out-of-plane ($z$)
and in-plane ($xy$) magnetizations (Shown in figure \ref{fig1}).

The system comes from an ordered state at low temperature to a
disordered state at high temperature. That behavior indicates an
order-disorder phase transition at $T_c \approx 0.55$. The in-plane
magnetization, $M_{xy}$, grows presenting a maximum close to the
order-disorder critical temperature $T_c$. However, the height of
the peak diminishes as $L$ grows, in a clear indicative that it is a
finite size artifice.

\begin {figure} [htbp]
\vspace{0.5cm}
\includegraphics [height=6.0cm,width=6.5cm] {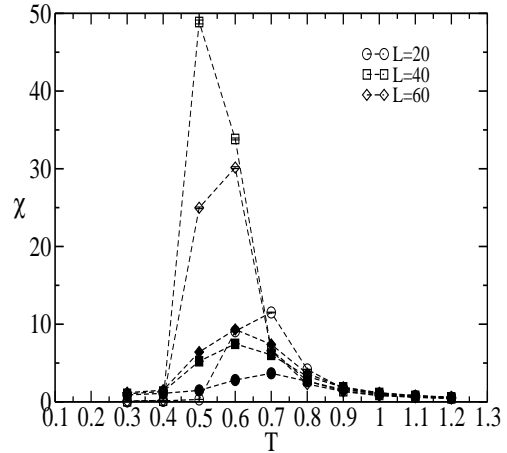}
\caption{Out-of-plane (open symbols) and in-plane (full symbols)
susceptibilities for $\delta =0.1$.} \label{fig2}
\end{figure}

The magnetic susceptibility is shown in figure \ref{fig2}. The
position of the maxima give us an estimate for  $T_c (\approx
0.55)$.

\begin{figure} [htbp]
\vspace{0.5cm}
\includegraphics [height=6.0cm,width=6.5cm] {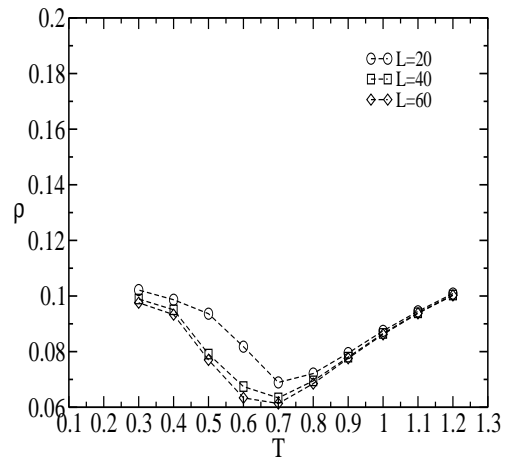}
\caption{Vortex density in the $xy$ plane for $\delta=0.1$.}
\label{fig3}
\end{figure}

We also measured the vortex density in the $xy$ plane as a function
of the temperature. Starting from the highest temperature, $T=1.2$,
the number of vortex decreases and reaches a minimum. Then it starts
to increase as the system is cooled down. This behavior is shown in
figure \ref{fig3} and the graphics indicate that the ground state of
the system has a significant number of vortices and anti-vortices in
the $xy$ plane. Apparently, the minimum of the vortex curve is
connected with the transition to in-plane magnetization, however, we
were not able to establish that connection.

\begin{figure} [htbp]
\vspace{0.5cm}
\includegraphics [height=6.0cm,width=6.5cm] {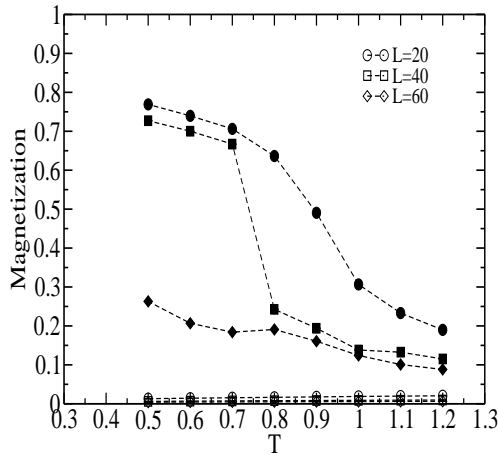}
\caption{Out-of-plane (open symbols) and in-plane (full symbols)
magnetization for $\delta =1.0$.} \label{fig4}
\end{figure}

\begin{figure} [htbp]
\vspace{0.5cm}
\includegraphics [height=6.0cm,width=6.5cm] {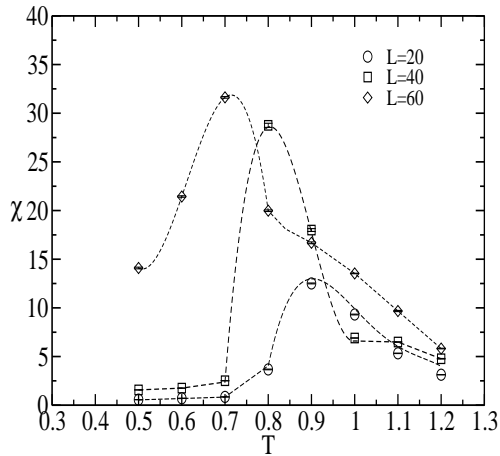}
\caption{In-plane susceptibility for $\delta =1.0$. The lines are
only a guide to the eye.} \label{fig5}
\end{figure}

For $\delta = 1.0$ the behavior of the in-plane and out-of-plane
magnetizations  (See figure \ref{fig4}), suggest that the ground
state is disordered in contrast to earlier works of Santamaria
\cite{santamaria} and Vedmedenko \cite{vedmedenko} that argue that
the ground state is for spins ordered in the $xy$ plane. A plot of
the susceptibility is shown in figure \ref{fig5} as a function of
temperature. Although some authors \cite{santamaria,vedmedenko}
concluded that this transition is of second order, the curves show
well defined maxima that do not seem to indicate any critical
behavior.

\begin{figure} [htbp]
\vspace{0.5cm}
\includegraphics [height=6.0cm,width=6.5cm]{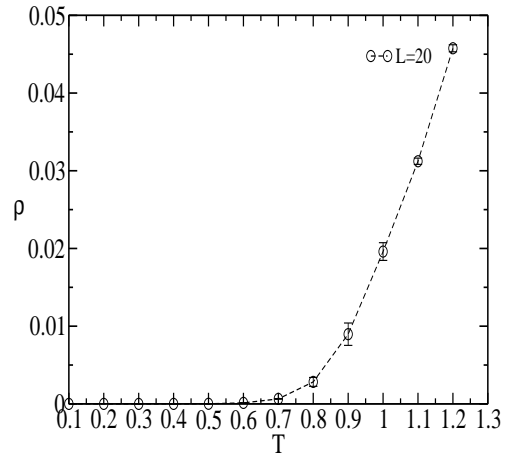}
\caption{Vortex density in the $xy$-plane for $\delta = 1.0$.}
\label{fig6} \vspace{-0.5cm}
\end{figure}

\begin{figure} [htbp]
\vspace{0.5cm}
\includegraphics [height=5.0cm,width=5.0cm]{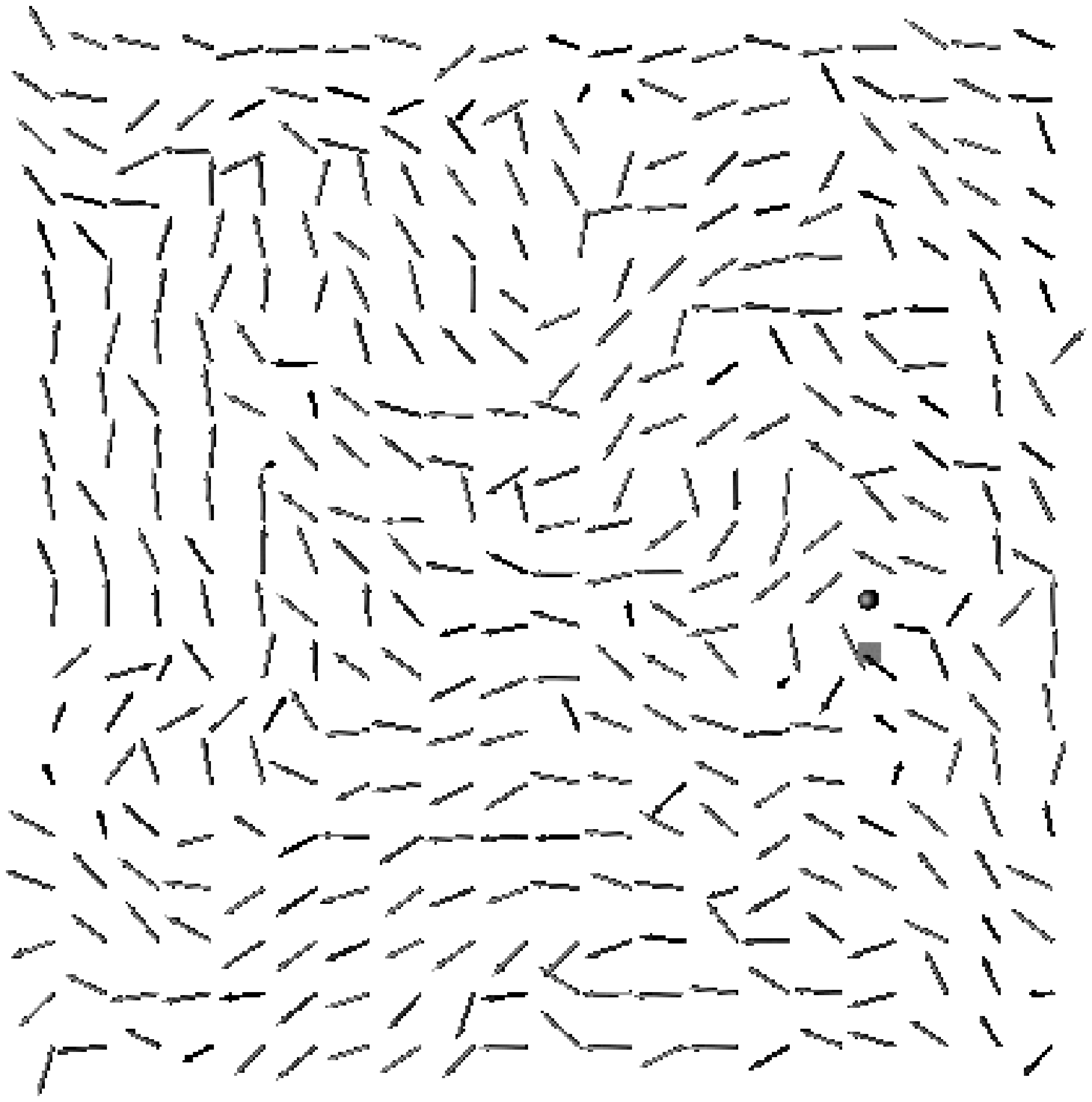}\hspace{-0.5cm}
\includegraphics [height=5.0cm,width=5.0cm]{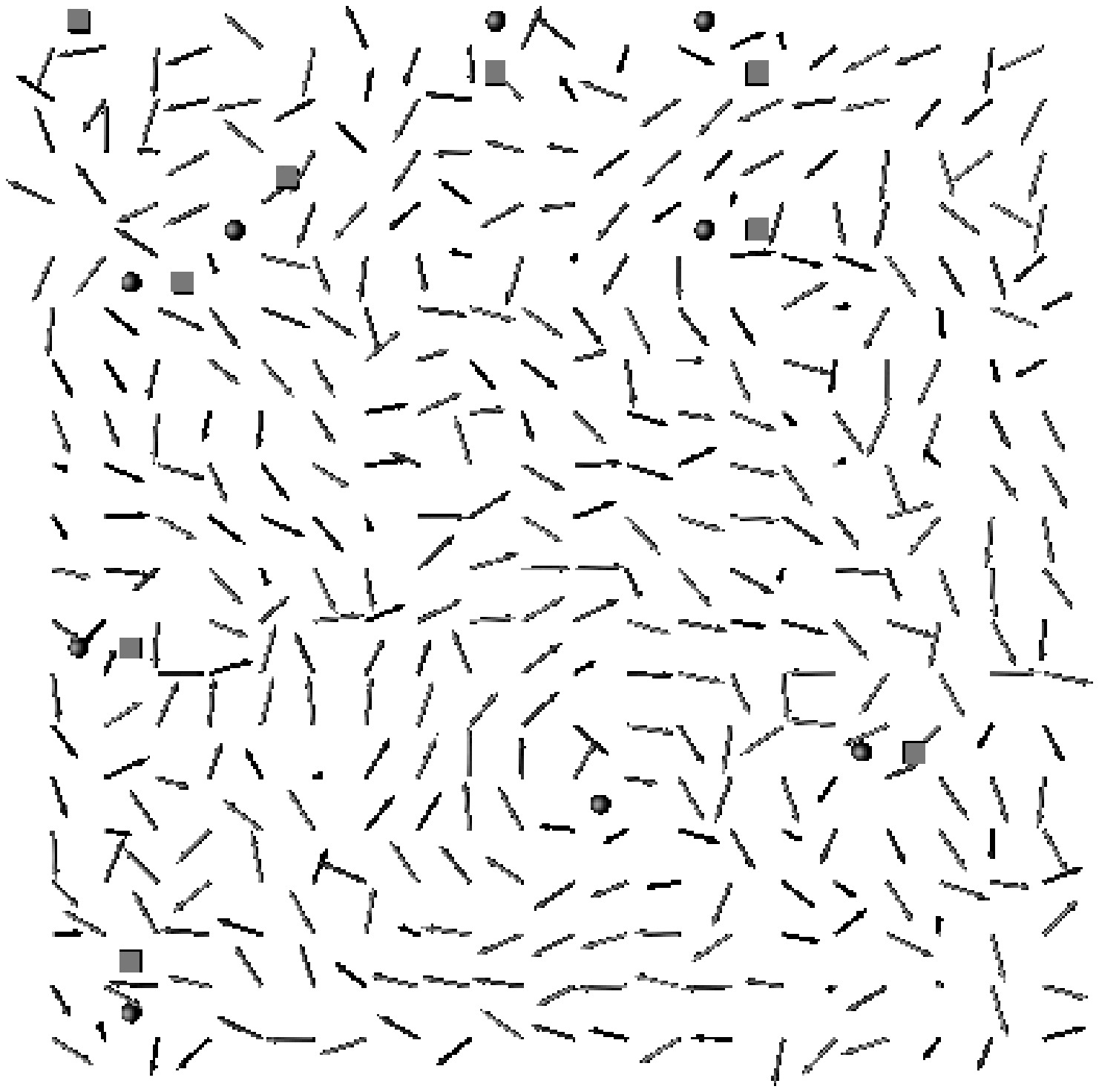}
\caption{Configurations of the system with $\delta= 1.0$ for $T=0.8$
e $T=1.2$. The vortices are indicated by spheres and the
anti-vortices by cubes.} \label{fig7} \vspace{-0.5cm}
\end{figure}

The vortex density curve in the $xy$ plane is shown in figure
\ref{fig6}. We see that the number of vortices increases
monotonically from zero as a function of temperature. As temperature
grows we observed that the spins in the lattice start to disorder,
so that pairs vortices-anti-vortices can unbind inducing a $BKT$
transition. However our results are not refined enough to decide
that. In figure \ref{fig7} we show two typical configurations for
$T=0.8$ and $1.2$ where the vortices are indicated by circles and
the anti-vortices by squares.

\begin{figure}[htbp]
\vspace{0.5cm}
\includegraphics [height=6.0cm,width=6.5cm]{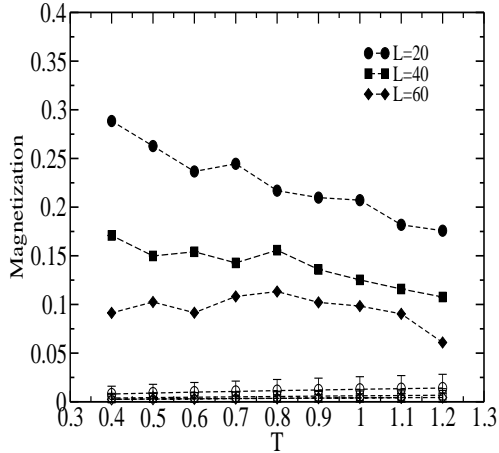}
\caption{Out-of-plane and in-plane magnetization (open and full
symbols) for $\delta = 9.0$.} \label{fig8}
\end{figure}

\begin{figure}[htbp]
\vspace{0.5cm}
\includegraphics [height=6.0cm,width=6.5cm]{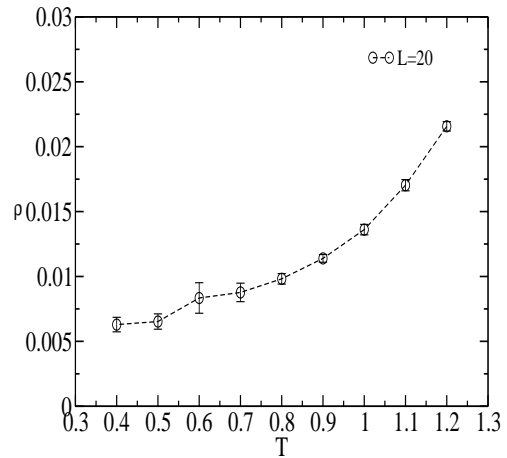}
\caption{Vortex density in the $xy$-plane for $\delta = 9.0$.}
\label{fig9}
\end{figure}

For systems with larger $\delta$, for example, $\delta = 9.0$, the
spins are preferentially in the $xy$ plane but it does not present
any magnetic ordering (See figure \ref{fig8}).The vortex density
curve is similar to the case where $\delta = 1.0$ (See
figure\ref{fig9}).

\section{Conclusion}
In summary, we investigated the Heisenberg spin model with exchange
$J$ and dipolar interactions $D$ and an anisotropic term $A$ for
different parameters $\delta=\frac{D}{A}$. For small $\delta$,
($0.1$), we observed that the vortex density has a minimum and is
non-zero for low temperatures. Apparently, this minimum is connected
with the order-disorder phase transition but this connection has to
be studied more carefully. For larger values of $\delta$ ($1.0$ and
$9.0$) the vortex density and the configurations of vortices in the
system led us to suspect of a phase transition of the BKT type
involving the unbinding of vortices-anti-vortices pairs. However our
results are not refined enough to decide that.


\end{document}